\documentclass[pre,reprint]{revtex4-1}

\usepackage[latin1]{inputenc}
\usepackage{amsmath}
\usepackage{graphicx}
\usepackage{nicefrac}
\usepackage{bbold}

\newcommand{\ket}[1]{\left| #1 \right>} 
\newcommand{\bra}[1]{\left< #1 \right|} 
\newcommand{\tr}{\text{tr}}

\begin{document}



\title{Macroscopic realism of quantum work fluctuations}

\author{Ralf Blattmann and Klaus M{\o}lmer}

\affiliation{Department of Physics and Astronomy,
             Aarhus University, DK-8000 Aarhus C, Denmark}


\date{\today}

\begin{abstract}
We study the fluctuations of the work performed on a driven quantum system, defined as the 
difference between subsequent measurements of energy eigenvalues. These work fluctuations are governed by statistical 
theorems with similar expressions in classical and quantum physics. In this article we show that 
we can distinguish quantum and classical work fluctuations, as the latter can be described by a
macrorealistic theory and hence obey Leggett-Garg inequalities. We show that these inequalities are violated by
quantum processes in a driven two-level system and in a harmonic oscillator subject to a squeezing transformation.
\end{abstract}


\maketitle


\section{Introduction}

The thermodynamics of quantum systems has become a rapidly expanding
field of research in the recent years \cite{Kosloff13,Goold16}.
One of its main lines of research has been triggered by the discovery of
fluctuation relations, especially the classical
work fluctuation theorems \cite{Jarzynski97,Crooks99}
and the derivation of their
quantum-mechanical counterparts \cite{Campisi11a}.
One prominent example of such a fluctuation theorem is the Jarzynski
relation \cite{Jarzynski97}

\begin{equation}
  \label{eq:Jarzynski}
    \langle e^{-\beta w}\rangle_w = e^{-\beta \Delta F},
\end{equation}
which relates the average work $w$ performed during
a non-equilibrium transformation with the free energy difference
$\Delta F$ between two thermal states at
inverse temperature $\beta = 1/k T$.
It has been shown that, if the work performed on a quantum system is defined
in a suitable way, the Jarzynski equality Eq.~\eqref{eq:Jarzynski}
holds for classical as well as for quantum systems \cite{Talkner07}. 
While there has been some debate what ``suitable'' means in this context
\cite{Campisi11a,Talkner16,Deffner16}, a widely accepted definition of
quantum work is given by the difference in the outcome 
of projective measurements of the Hamiltonian operator at different times.
With this definition the quantum work becomes, in general,
a fluctuating quantity, similar to the fluctuating work
in classical non-equilibrium thermodynamics.
While the (non-equilibrium) work is characterized
by a (classical) probability density in both cases, the origin of the randomness
is quite different:
In classical physics energies have definite values, and work fluctuations stem from the random exchange of energy with the particles in the surrounding heat bath,
while in quantum mechanics, they originate from quantum uncertainty and the resulting fundamental 
randomness of the outcome of measurement, i.e., Born's rule and the projection postulate.
In this work, we address the question: ``\textit{How quantum is quantum work ?}'', by investigating to what extent the different origin of quantum and 
classical work fluctuation have measurable consequences.
\\
Our approach to the problem will be to consider processes where the energy of the system is measured multiple times, see Fig.1, and where we can hence study temporal 
correlations in the work fluctuations. If the work behaves classically, it should be describable as a macroscopic, realistic variable, which is measurable in a non-invasive 
manner, and hence its correlations should obey the Leggett-Garg inequalities \cite{Leggett85, Emary14}.
Leggett-Garg inequalities have been used to analyze quantum effects in thermodynamics
processes and heat engines \cite{Friedenberger15}. While they only apply for correlation functions of dichotomic variables  in their original
form, entropic Leggett-Garg inequalities have been derived for correlations of more general variables.
\\
The article is structured as follows: In Sec.~\ref{sec:Work}
we introduce the definition of quantum work and its probability
distribution. In Sec.~\ref{sec:Leggett}
we recall the dichotomic and entropic Leggett-Garg inequalities and we discuss their application to the work done on a quantum system.
In Sec.~\ref{sec:TLS} and Sec.~\ref{sec:Squeeze}
we investigate whether the inequalities are obeyed or violated for a driven two-level
system and a squeezed harmonic oscillator, respectively.
In Sec.~\ref{sec:Conclusion} we discuss the consequences and possible applications of our results.

\section{Quantum Work and its probability distribution}
\label{sec:Work}

Consider an isolated quantum system with a time-dependent
Hamiltonian $H_t=H(\lambda_t)$, where $\lambda_t$
is a varying  control parameter. The state of the system obeys the
Liouville-von Neumann equation $i\hbar\dot{\rho}_t = [H_t,\rho_t ]$, and, in general, 
work will be performed on it, i.e. energy will be injected into (or removed from) the system.
In order to determine the work performed during a given process, 
(quantum as well as classically), one measures the energy of the system
before and after the process. While in classical systems such
measurements are unproblematic, the measurement on a quantum system
will in general have random outcomes and it will change the state of the system. 
\begin{figure}
    \includegraphics[scale=1]{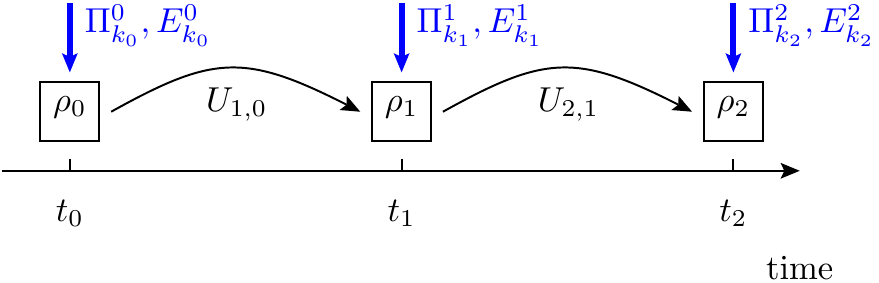}
    \caption{Illustration of our work measurement process. The system is prepared in an initial thermal state, and a measurement of its energy
     projects it into an energy eigenstate $|k_0\rangle$ of the Hamiltonian $H_{t_0}$ at $t=t_0$. The 
              Hamiltonian leads to a time evolution of the quantum state between $t_0$ and the later time $t_1$, which can be expressed as a unitary operator
              $U_{1,0}$, which governs the expansion of the state on the energy eigenstates of $H_{t_1}$ and causes the stochastic nature of the energy measurements at $t1$. 
              The work is defined as the difference $E^1_{k_1}-E^0_{k_0}$ between the system eigenenergies measured,
              and by evolving the system further and measuring the energy at the later time $t_2$,  
              we can study correlations between the work done during the two evolution processes.}
  \label{fig:Model}
\end{figure}
If one wants to measure the work performed on the system between the times $t_0$ and
$t_1$ one has to probe the system energy at the beginning and the end 
of the driving. This will yield one of the eigenvalue $E^0_{k_0}$ of $H_{t_0}$
with a probability $p_{k_0}= \tr[ \Pi^0_{k_0} \rho_{0}]$, and, subsequently, an eigenvalue $E^1_{k_1}$ of $H_{t_1}$, 
with a probability $p_{k_1}= \tr[ \Pi^1_{k_1} \rho_{1}]$, where we have introduced the projection operators on the energy eigenstates,
$\Pi^\alpha_{k_\alpha}  =   \ket{k_\alpha}   \bra{k_\alpha}$ of $H_t$ at $t=t_\alpha,\ \alpha=0,1$. The state at $t_1$ is $\rho_1= U_{1,0}\rho U^{\dagger}_{1,0}$, 
where $U_{1,0}=\mathcal{T}\exp[-i\int_{t_0}^{t_1} H_t dt]$  is the time evolution operator
from time $t_0$ to $t_1$ ($\mathcal{T}$ denotes the time ordering operator) and $\rho=\Pi^0_{k_0} $ is the state conditioned on the outcome of the 
first measurement at $t=t_0$. Subsequent time evolution of the quantum state and measurements are described by the same formalism, cf., Fig. 1.

One obtains the work by merely subtracting the measured energies, 
\begin{equation}
  \label{def:work}
     W_{k_1,k_0}  = E^1_{k_1} - E^0_{k_0},
\end{equation}
and since it depends on the random outcome of projective energy measurements
it is an inherently fluctuating quantity.
Its probability distribution is given by \cite{Campisi11a}
\begin{equation}
  \label{eq:wpd}
 p(w) =\sum_{k_0,k_1} \delta(w - W_{k_1,k_0} ) p_{k_1,k_0},
\end{equation}
where, according to the above arguments,
\begin{equation}
  \label{eq:joint_E}
    p_{k_1,k_0} = \tr[\Pi^1_{k_1}U_{1,0}
        \Pi^0_{k_0} \rho_{0}  \Pi^0_{k_0} U^{\dagger}_{1,{0}} ].
\end{equation}
is the joint probability distribution for measuring $E^0_{k_0}$ at $t_0$ and $E^1_{k_1}$
at $t_1$.
\\
Assuming that the system is initially in a thermal state
$\rho_0= \exp(-\beta H_{0})/Z_0$ with $Z_0= \tr[-\beta H_{0}]$,
it is rather straightforward to use the characteristic function \cite{Talkner07} 
for the work probability distribution Eq.~\eqref{eq:wpd} and verify that
the quantum Jarzynski equality Eq.~\eqref{eq:Jarzynski} holds,
where $\langle \dots \rangle_w= \int \dots p(w) dw $
and $\Delta F = F_{1}-F_{0}$ with the free energy  $F_t=-(1/\beta) \ln (Z_t)$.

Instead of measuring the energy only in the beginning and the end of a process
one might perform several energy measurements at different times,
see Fig.~\ref{fig:Model}.
Probing the system energy at three instants of time, $t_0,t_1,t_2$
then leads to the joint probability distribution
%
%
%
%
\begin{align}
  \label{eq:joint_3}
   p_{k_2,k_1,k_0}
   = \tr[\Pi^2_{k_2}
           U_{2,1} \Pi^1_{k_1} U_{1,{0}}
            \Pi^0_{k_0} \rho_0  \Pi^0_{k_0} U_{1,{0}}^{\dagger}
            \Pi^1_{k_1}  U^{\dagger}_{2,1} ].
\end{align}
By summing over indices one obtains, e.g., $p_{k_1,k_0}$ and $p_{k_2,k_1}$ from (\ref{eq:joint_3}). 
Note that, since in general $[\Pi^j_{k_j},U_{i,j}]\neq 0$,
the distribution arising from summing over the middle index $k_1$
is not equivalent to the distribution $p_{k_2,k_0}$ without measurement
at $t_1$.
\\
From Eq.~\eqref{eq:joint_3} we define the joint probability
%
%
\begin{align}
  \label{eq:work_corr}
   p(w_{1}, w_{2})
   = \sum_{k_0,k_1,k_2}& \delta(w_{1}- W_{k_0,k_{1}})  \delta(w_{2}- W_{k_1,k_{2}})\\
     &\times   p_{k_2,k_1, k_0}  \notag
\end{align}
to perform the work $w_{1}$ between $t_0$ and $t_1$, and the work $w_{2}$
between $t_1$ and $t_2$.
Note that its marginal distribution $p(w_1)=\int dw_2 p(w_1,w_2)$
is equivalent to Eq.~\eqref{eq:wpd} and hence fulfills the fluctuation relations, while
the marginal $p(w_2)=\int dw_1p(w_1,w_2)$
will in general not fulfill such relations because the
system is not in an equilibrium state at $t_1$.

Using Eq.~\eqref{eq:work_corr} the probability
for the total work $w_\text{tot} = w_1+w_2$ yields
%
%
%
%
\begin{align}
p(w_\text{tot}) =& \int dw_{1} dw_{2} \delta(w_\text{tot}-[w_1+w_2])
                  p(w_{1}, w_{2}) \notag\\
= & \sum_{k_0,k_1,k_2} \delta(w_\text{tot} - W_{k_2,k_0}) p_{k_2,k_1, k_0},
\end{align}
i.e., it depends, as one would expect, only on the difference between the first
and the last energy measurements. 
While the intermediate measurements will, in general,
influence the final energy measurement \cite{Campisi11b}, one can prove that the Jarzynski relation \eqref{eq:Jarzynski} 
still holds for the total work \cite{Campisi10a,Campisi11b}.

The measurement back action and, in particular, the destruction of quantum mechanical coherence by the middle measurement
(the first measurement acts on a thermal state with already vanishing coherences), presents a fundamental difference between the definition of 
work in quantum and classical contexts. One may, indeed, include non-invasiveness as a desired property of the definition of work, but that
turns out to be incompatible with its relationship with the average energy of the system \cite{Acin16}. 

In this article, we shall instead retain the generally accepted definition of work, and address its invasiveness in a more quantitative manner. 
To this end, we shall appeal to the Leggett-Garg inequalities \cite{Leggett85,Emary14}, which precisely concern correlations between measurements 
performed at different times on a quantum system. We note that Leggett-Garg inequalities have also been applied to characterize the quantumness of 
a quantum heat engine through the correlation between the working system observables at different times \cite{Friedenberger15}.

\section{Legget-Garg inequalities for work measurements}
\label{sec:Leggett}

Assuming macroscopic realism and noninvasive measurability of a dichotomic variable 
that is measured at different times, $t_i$ with output values $Q_i=\pm 1$, 
Leggett and Garg derived the inequality \cite{Leggett85,Emary14}
\begin{equation}
 \label{eq:LG}
    C_{21} + C_{32} - C_{31} \leq 1.
\end{equation}
for the two-time correlation functions $C_{ij}= \langle Q_i Q_j\rangle$. While Eq.~\eqref{eq:LG} is obeyed for
classical dynamics, measurements on a quantum systems may violate the Leggett-Garg inequality \cite{Palacios-Laloy2010,Knee2012}. 
This violation is readily understood as a consequence of the measurement back action, which is absent in the classical case. In the 
present work, we shall use the Leggett-Garg inequality to assess how the definition of quantum work as the result of projective energy 
measurements necessarily implies a quantitative difference between the fluctuations of quantum and classical work.  
Note that  Eq.~\eqref{eq:LG} is not associated with the absolute magnitude of energy and work measurements, but only the statistical correlations of
the variables $Q_i=\pm 1$, which we can associate with the projective measurements on two different eigenstates. 

For systems with more eigenstates, 
the measurement outcome $Q_i$ at time $t_i$ may attain more than two values $\{q_i\}$, and  an
alternative, entropic Leggett-Garg inequality has been derived for the correlations between such multi-valued measurement outcomes \cite{UshaDevi13},
\begin{equation}
 \label{eq:entropic_LG}
  H(Q_{2}|Q_{0}) \leq  H(Q_{2}|Q_{1}) + H(Q_{1}|Q_{0}).
\end{equation}
Here, $ H(Q_{j}|Q_{i})=- \sum_{q_j,q_i} p(q_i) p(q_j|q_i) \log p(q_j|q_i)$
is the (classical) conditional entropy, where $p(q_j|q_i)$ is the conditional probability for outcome $q_j$ given the earlier outcome $q_i$ (occurring with probability $p(q_i)$).
\\
We shall now apply the Leggett-Garg and entropic Leggett-Garg inequalities
to the correlations Eq.~\eqref{eq:joint_E} of energy measurements. 
These correlations reflect how the definition of work is affected by measurement back-action effects and, hence, to what extent 
the underlying thermodynamic transformation can be modelled  as a classical process.
\\
The mean conditional entropy
for energy measurements at two instants of time is given by
\begin{align}
  \label{eq:en_entropy}
    H(E^j|E^i)  =& - \sum_{k_i,k_j} p_{k_j,k_i} \log p_{k_j|k_i},
\end{align}
where the conditional probability, $p_{k_j|k_i}$, is the quantum mechanical transition probability  
$\ket{k_i}\rightarrow \ket{k_j}$, between the eigenstates of the Hamiltonian at times $t_i$ and $t_j$, as governed by the 
unitary time evolution operator $U_{j,i}$, defined above.
\\
For a slowly varying Hamiltonian, the system will adiabatically follow the time dependent eigenstate in which it is prepared by the first measurement, and hence 
$p_{k_1|k_0}=\delta_{k_1,k_0}$ (note the eigenenergies may generally differ and a definite, 
nonvanishing amount of work is hence done on the system). Also, during the subsequent evolution and measurement, 
the system follows the same ($k^{th}$) eigenstate, and due to the definite outcomes, $H(E^j|E^i)=0$ and the
entropic Leggett-Garg equality Eq.~\eqref{eq:entropic_LG} for energy measurements is trivially fulfilled. In the following sections, 
we shall hence consider systems that do not evolve adiabatically.
\\
Noting that the probability distribution for the work done on the system is governed by the joint probability of the two pertaining energy measurements, 
$w_{k_j,k_i}=E^j_{k_j}-E^i_{k_i}$,
$p(w_{k_j,k_i})=p(k_j,k_i)=p(k_j|k_i)p(k_i)$, we shall introduce the corresponding entropy $H(w_{ji}) =H(E^j,E^i) =-\sum_{k_i,k_j} p_{k_j,k_i} \log  p_{k_j,k_i}$.
Using the identity between conditional and joint entropies \cite{Cover91}, $H(E^j|E^i) = H(E^j,E^i) - H(E^i)$,
where $H(E^i_{k_i}) = -\sum_{k_i} p_{k_i} \log  p_{k_i}$
is the entropy of the distribution $p_{k_i}$, we can rewrite
the entropic Leggett-Garg inequality (\ref{eq:entropic_LG}), as a relation for the work distributions
\begin{align}
  \label{eq:work_LG}
  \mathcal{H}(w_{20}) \leq \mathcal{H}(w_{21}) + \mathcal{H}(w_{10}) - H(E^1).
\end{align}
Here, $\mathcal{H}(w_{ij}) = -\sum_{ p(w_{ij})\neq 0 } p(w_{ij}) \log  p(w_{ij})$
is the entropy of the work distribution, which is discrete
if the corresponding energy spectrum is discrete.
Note, that Eq.~\eqref{eq:work_LG} does not only
depend on the entropy of work distributions but also
on the entropy of the middle energy measurement
$H(E^1)$. Disregarding this term leads
to an inequality that is more easily fulfilled, and which reflects that
in a classical system the entropy of the
work distribution, without the middle measurement
at $t_1$ is always smaller than with this measurement taking place,
because classical measurements do not decrease the
information \cite{UshaDevi13}. Note that it is easier to observe a violation of the Legget-Garg inequality if the 
entropy of the middle energy measurement
$H(E_1)$ is retained in Eq.~\eqref{eq:work_LG}.

As a side remark, we note that to go from Eq.~\eqref{eq:en_entropy} to Eq.~\eqref{eq:work_LG}
we assume that the the work distribution has no ``degeneracies'',
i.e.~there are no $k_j,k_i$ and $k'_j,k'_i$ with $W_{k_j,k_i}=W_{k'_j,k'_i}$.
While such degeneracies may be easy to avoid, they can also be accounted for by using 
the grouping formula for the Shannon entropy \cite{Cover91} where joint probabilities with
$W_{k_j,k_i}=W_{k'_j,k'_i}$ are grouped together.
If ${\bf p} =\{p_1,\dots,p_n\}$ is a probability distribution
and ${\bf q} =\{q_1,\dots,q_m\}$ with $q_j=\sum_{i\in I_j} p_i$
is another distribution formed by ``grouping'' the
probabilities $p_i$ which correspond to a subset of events with indices $I_j$
the Shannon entropy yields
\begin{align}
    \label{eq:grouping}
    H({\bf q})& = H({\bf p})
                         - \sum_j q_j H\Bigl( \bigl\{\frac{p_i}{q_j}
                            | i \in I_j \bigr \} \Bigr )   \notag  \\
                        & = H({\bf p}) - \bar{H}({\bf q}).
\end{align}
Hence, the grouping reduces the Shannon entropy by an amount given
by weighted entropies of subset probabilities.

\section{Violation of the conventional and the entropic Leggett-Garg inequalities for a two-level system}
\label{sec:TLS}

We recall that the work is defined through projective measurements in the eigenstate basis of the time dependent Hamiltonian.
One should hence determine the time evolution operator by solving the Schr\"odinger equation, and subsequently express it as a 
transformation between the eigenstates at different times. Since both the time evolution of the state of the system and the time dependence of the eigenstates 
are governed by unitary operations, the evolution with respect to the adiabatic  basis is also unitary, and can, e.g., for the evolution between the first 
and the middle measurement, be expressed as
\begin{equation}
   \label{eq:U_mat}
  \bar{U}_{1,0} = \begin{pmatrix}
                                   e^{\frac{i}{2}(\alpha+\beta)}\cos \frac{\theta}{2}
                                   & e^{\frac{i}{2}(\alpha-\beta)}\sin \frac{\theta}{2}\\
                                  - e^{\frac{i}{2}(\alpha-\beta)}\sin \frac{\theta}{2}
                                   & e^{\frac{-i}{2}(\alpha+\beta)}\cos \frac{\theta}{2}
                   \end{pmatrix},
\end{equation}
with angles $\alpha$, $\beta$ and $\theta$.
Note that this matrix pertains to the time dependent adiabatic basis, and when the mixing angle $\theta$ is small, it represents the case of 
adiabatic evolution (slowly varying Hamiltonian). In this limit, the system adiabatically follows the energy eigenstates of the system, 
the measurements are fully correlated and the entropies vanish. Rather than specifying a time dependent Hamiltonian, it is convenient and it represents 
no loss of generality, to represent the dynamics by the transformation matrix with respect to the time-dependent energy eigenstates
$\bar{U}_{1,0}$ and the similarly defined $\bar{U}_{2,1}$. The matrix 
elements of these matrices then directly yield the joint probabilities in Eq.~\eqref{eq:joint_E}.
\begin{equation}
  \label{eq:trafo_join}
  p_{k_1,k_0}  = \tr[\Pi^1_{k_0} \bar{U}_{1,0} \Pi^0_{k_0} \rho_0 \Pi^0_{k_0}  \bar{U}^{\dagger}_{1,0}]
\end{equation}
For simplicity of analysis, we set $\alpha=\beta=0$, so that $\bar{U}_{1,0}$ becomes a real
rotation matrix. The angle $\theta$ in Eq.~\eqref{eq:U_mat} controls
how much population is transfered between the eigenstates at different times, and, for simplicity, 
we shall assume that $\bar{U}_{2,1}=\bar{U}_{1,0}$.
\\
\textit{Mutatis mutandis}, we can obtain the joint probabilities 
$p_{k_2,k_0}$ and $p_{k_2,k_1}$, so that we can study the 
violation Eq.~\eqref{eq:LG}, where the dichotomic variable takes the values $\pm 1$
in the ground and excited state, and of Eq.~\eqref{eq:work_LG}. 

\begin{figure}
    \includegraphics[scale=0.55]{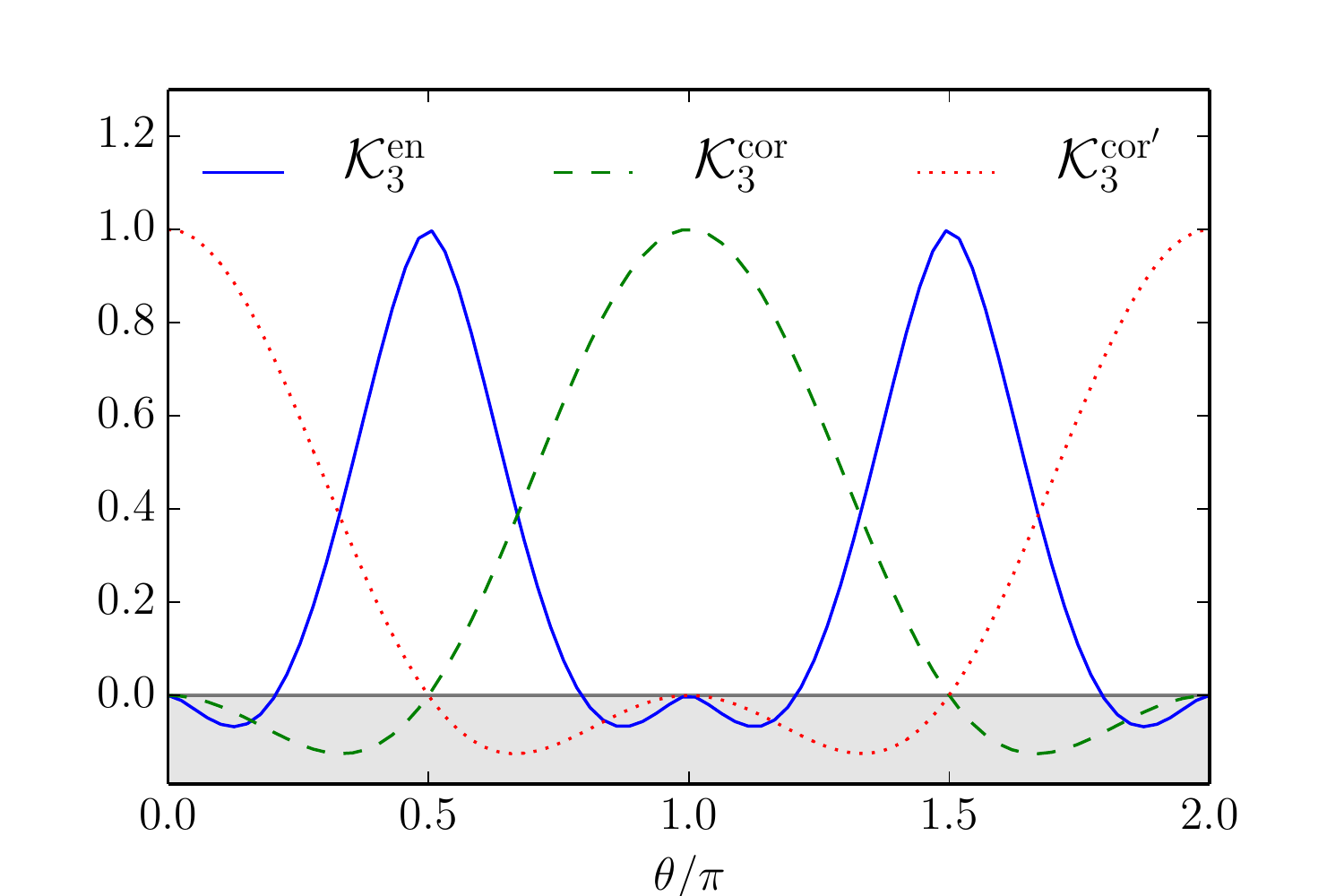}
    \caption{The Leggett-Garg parameters Eqs.~\eqref{eq:LG_param_cor} and
             \eqref{eq:LG_param_en} as a function of the angle $\theta$.
              Negative values (the gray area) imply a violation of the
              corresponding inequality.}
     \label{fig:LG_TLS}
\end{figure}

In order to quantify the violation of the equations
Eq.~\eqref{eq:LG} and, respectively, Eq.~\eqref{eq:work_LG}
we define the Leggett-Garg parameters
\begin{equation}
  \label{eq:LG_param_cor}
     \mathcal{K}^\text{cor}_{3}=\frac{1}{4}(1- C_{01} - C_{02} + C_{12})
\end{equation}
and,
\begin{equation}
  \label{eq:LG_param_en}
     \mathcal{K}^\text{en}_{3}=\frac{1}{2}\bigl(\mathcal{H}(w_{21}) + \mathcal{H}(w_{10})
                       - \mathcal{H}(w_{20}) - H(E_1) \bigr),
\end{equation}
where negative values of the parameters are a signature
of non-classical behavior.
In Fig.~\ref{fig:LG_TLS}, we plot $\mathcal{K}^\text{cor}_{3}$
and $\mathcal{K}^\text{en}_{3}$ as
functions of the angle $\theta$. Moreover, we plot $\mathcal{K}^\text{cor'}_{3}$
which follows from $\mathcal{K}^\text{cor}_{3}$ by redefining $Q_1 \rightarrow -Q_1$ \cite{Emary14}. 
As initial state we chose a thermal state with $\beta=1/\Delta E$, where $\Delta E$ is the energy splitting of the ground and excited state.
Actually, for a two-level system the violation of the
Leggett-Garg inequality does not depend on the initial state and temperature, as the matrix $\bar{U}_{1,0}$ yields the same probability to obtain the 
same and the opposite eigenstates in the middle measurement for both initial outcomes.
\\
Fig.~\ref{fig:LG_TLS} shows that for small finite $\theta$, all
curves are in the grey area where the Leggett Garg inequality is violated.
They, however, differ considerably. While either $\mathcal{K}^\text{cor}_{3}$ or $\mathcal{K}^\text{cor'}_{3}$
violate the conditions for macroscopic realism for all angles except for multiples of $\pi/2$,
non-classical correlations between measurements are not always revealed
by the entropic Leggett-Garg inequality.

\section{Violation of the entropic Leggett-Garg inequality for a squeezed harmonic oscillator}
\label{sec:Squeeze}

In this section, we study the entropic Leggett-Garg inequality
for a harmonic oscillator. The quantum harmonic oscillator is in many aspects
well described by classical physics, e.g., the evolution of the continuous position and momentum operators solve the same coupled
linear equations as the classical coordinates. The work done on a harmonically trapped particles 
has been studied quite extensively in both the classical and the quantum case \cite{Deffner08,Monnai10,Horowitz12,Campisi13}.

The harmonic oscillator
is described by the Hamiltonian
\begin{equation}
 H = \frac{p^{2}}{2m} + \frac{1}{2}m\omega^2 x^2
 = \hbar\omega\Bigl( a^{\dagger}a + \frac{1}{2}\Bigr),
\end{equation}
with $a=\sqrt{m\omega/2\hbar} \bigl(x + (i/m\omega) p \bigr)$
and $[a, a^{\dagger}]=1$.
Driving the system with an arbitrary time-dependent potential strength $\omega_t$ will maintain the quadratic form of the Hamiltonian, and hence result in linear coupled equations 
for the position and momentum operators or, equivalently, for the raising and lowering operators. Their time dependence in the Heisenberg picture can hence be 
represented by a Bogoliubov (squeezing) transformation \cite{Bogoljubov58,Mollow67},%
\begin{equation} \label{eq:bogo}
 a(\tau)= U^{\dagger}_{\tau}\ a\ U_{\tau} = \mu(\tau) a + \nu(\tau) a^{\dagger},
\end{equation}
where $\mu(\tau)$ and $\nu(\tau)$ are complex functions depending
on details of the driving and $U_\tau$ is the corresponding time evolution operator.

As in the case of the two-level system, we shall express the evolution with respect to the operators defining the energy measurements, $H_t=\hbar\omega_t\Bigl( a_t^{\dagger}a_t + \frac{1}{2}\Bigr)$. These are the adiabatically evolved operators, and they are also given by a Bogoliubov transformation. Hence, without loss of generality, we can also represent the transformation of the quantum state, expressed in terms of the raising and lowering operators
pertaining to the time dependent Hamiltonian as a Bogoliubov, or squeezing, transform,
\begin{equation}
   \label{eq:squeeze_op}
   \bar{U}_{10}^{\dagger}\ a_{t_0} \bar{U}_{10} = \cosh r a_{t_1} + \sinh r e^{-i\phi} a_{t_1}^{\dagger}.
\end{equation}

For simplicity, we  omit the phase $\phi$, and using Eq.~\eqref{eq:squeeze_op} as the propagator between two energy measurements,
the joint probability distribution \eqref{eq:trafo_join} yields
\begin{align}
   \label{eq:squeeze_join}
    p_{k_1,k_0} &= \tr \Bigl[\Pi^1_{k_1} \bar{U}_{10} \Pi^0_{k_0} \rho_0 \Pi^0_{k_0}  \bar{U}_{10}^{\dagger} \Bigr] \\
            &= G^2_{k_1,k_0}(r) \rho_{k_0,k_0},
\end{align}
with $\rho_{nn}=\bra{n}\rho_0\ket{n}$ and $G_{mn}(r)=\bra{m}U_r\ket{n}$.
The transition matrix elements for squeezed number states $G_{mn}(r)$ are provided in the Appendix \ref{app:anal} following
analytical results in \cite{Satya85,Nieto97}.
With the above preparation, we are ready to address the entropic Leggett-Garg inequality ~\eqref{eq:work_LG},
where the time evolution between $t_0$ and $t_1$ and between $t_1$ and $t_2$ are both governed by (\ref{eq:squeeze_op}), but possibly with two different arguments $r_1$ and $r_2$. 
\begin{figure}
    \includegraphics[scale=0.55]{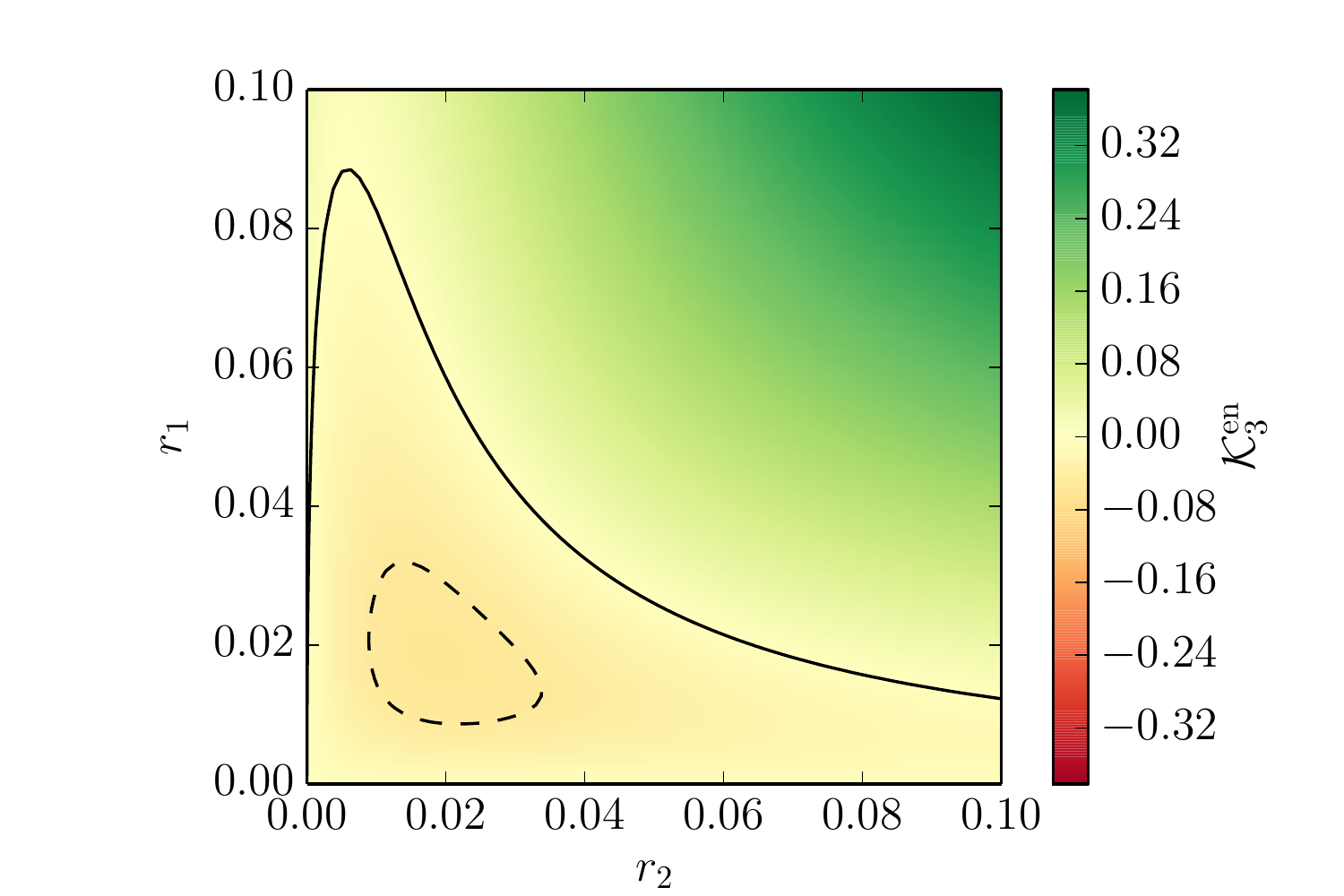}
    \caption{Violation of Eq.~\eqref{eq:work_LG} for a squeezed harmonic oscillator as
             a function of the squeezing parameter $z$ (between $t_0$ and $t_1$)
             and $\bar{z}$ (between $t_0$ and $t_1$). The solid contour encloses the area
             where $\mathcal{K}_3^\text{en}$ gets negative, i.e. where the inequality
             gets violated. The dashed contour denotes $\mathcal{K}_3^\text{en}=-0.05$.
             As initial state we assume a thermal state with $\beta= 0.1(\hbar \omega_0)^{-1}$}
     \label{fig:LG_squeeze_2d}
\end{figure}
%
\\
Plotting the entropic Leggett-Garg parameter $\mathcal{K}_3^\text{en}$
as a function of $r_1$ and $r_2$ for a thermal initial state
with $\beta=0.1(\hbar \omega_0)^{-1}$ in Fig.~\ref{fig:LG_squeeze_2d} reveals
that also for the harmonic oscillator, Eq.~\eqref{eq:work_LG}
can be violated and the quantum work obeys non-classical
statistics.
Interestingly, even a vanishing small amount of squeezing is enough
to violate the Leggett-Garg inequality. Increasing the squeezing
strength increases the violation until a maximal violation is obtained
at $r_1=r_2 \approx0.02$.
Further increase of the squeezing parameter leads to a less
pronounced violation and finally 
$\mathcal{K}^\text{en}_3$ turns positive indicating
that the works statistics can no longer be distinguished 
from that of a classical process.
This can be understood as a consequence of the fact that
the squeezing of thermal states and number states
generally broadens the number distribution, turning
sub-Poissonian into super-Poissonian statistics \cite{Kim89}.
Note also that for strong squeezing there is an asymmetry between $r_1$ (the squeezing between
$t_0$ and $t_1$) and $r_2$ (the squeezing between
$t_1$ and $t_2$).
In this regime, too much squeezing in the second interval prevents the violation of the Leggett-Garg inequality.
\begin{figure}
    \includegraphics[scale=0.55]{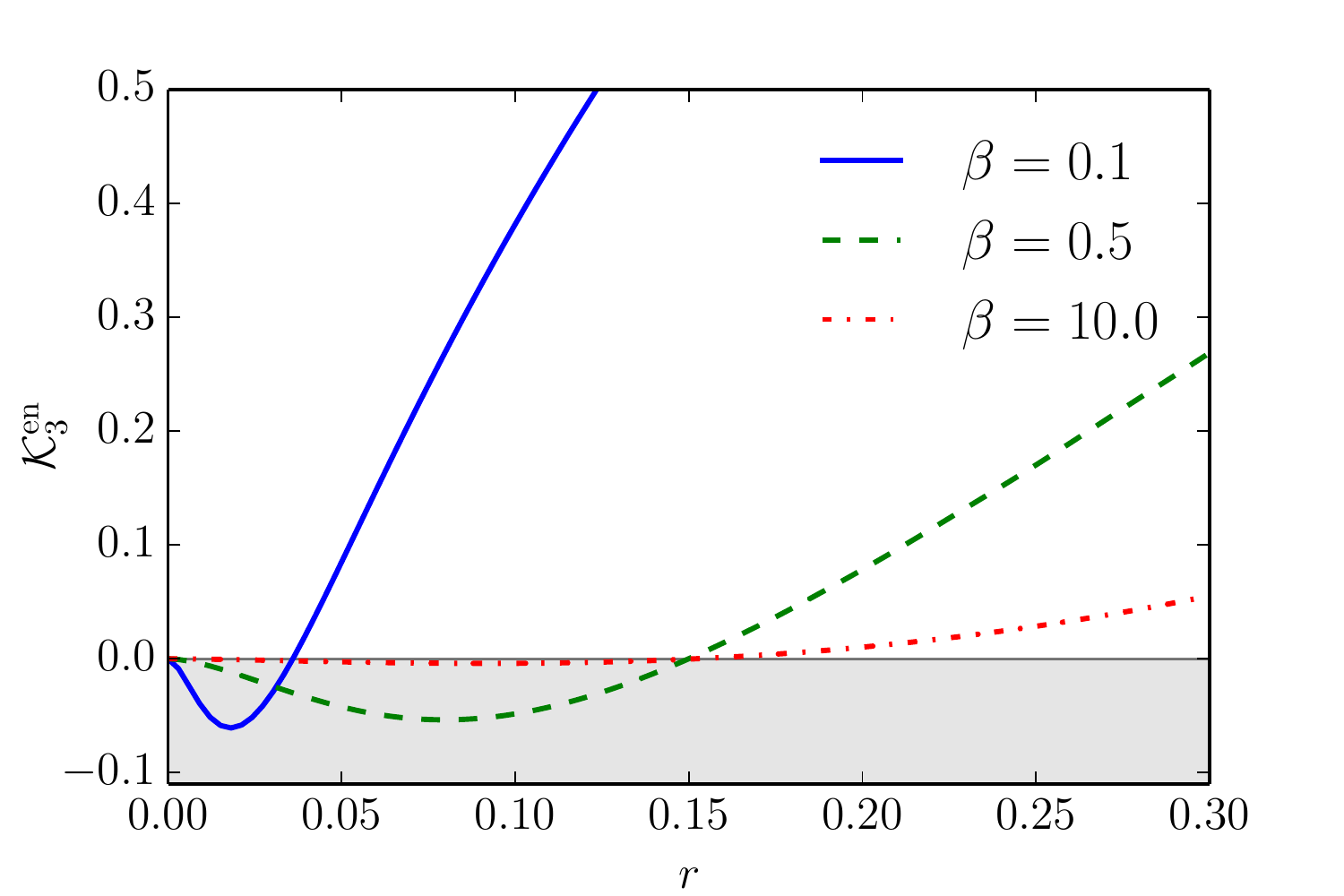}
    \caption{The entropic Leggett-Garg parameter for a squeezed harmonic oscillator
             as a function of the squeezing parameter and different
             thermal initial states
             with inverse temperatures $\beta$ expressed in multiples of $1/\hbar\omega_0$. }
     \label{fig:LG_squeeze}
\end{figure}
\\
It is an interesting question how the violation depends on the temperature of the thermal initial state.
We recall that for the two-level system, the outcome correlations are independent of the outcome of the first measurement, 
and hence of the initial state. This is different for the oscillator, since the probability for measuring high energy outcomes in the first measurement  
depends on the initial temperature and do affect the subsequent correlations. 
In Fig.~\ref{fig:LG_squeeze}, we plot the $\mathcal{K}^\text{en}_3$
for different values of inverse temperatures $\beta$
as a function of the squeezing parameter $r$, where we set $r_1=r_2$.
Interestingly, the violation for larger $\beta$, i.e.~lower
temperature, is less pronounced than for small $\beta$.
Hence, the quantum work statistics appears more non-classical for
higher initial temperatures. This can also be seen in the
upper plot of Fig.~\ref{fig:LG_squeeze_beta}, where we plot
the minimal value of $\mathcal{K}^{en}_3$ as a function of $\beta$.
This can be understood by the fact that for higher temperatures
the system is more likely to start in a higher number state after the 
first measurement which is more strongly affected by the squeezing.
\\
Fig.~\ref{fig:LG_squeeze} also reveals that for increasing $\beta$
the values for the squeezing parameter $r_1=r_2$, where the maximal violation occurs
increases. This is studied more generally
in Fig.~\ref{fig:LG_squeeze_beta}, where we plot the value of $r$ leading to maximal
Leggett-Garg violation as function of $\beta$.
We observe a non-monotonus dependency with a maximum around
$\beta=1$ and approach towards a constant level for large $\beta$.
\begin{figure}
    \includegraphics[scale=0.55]{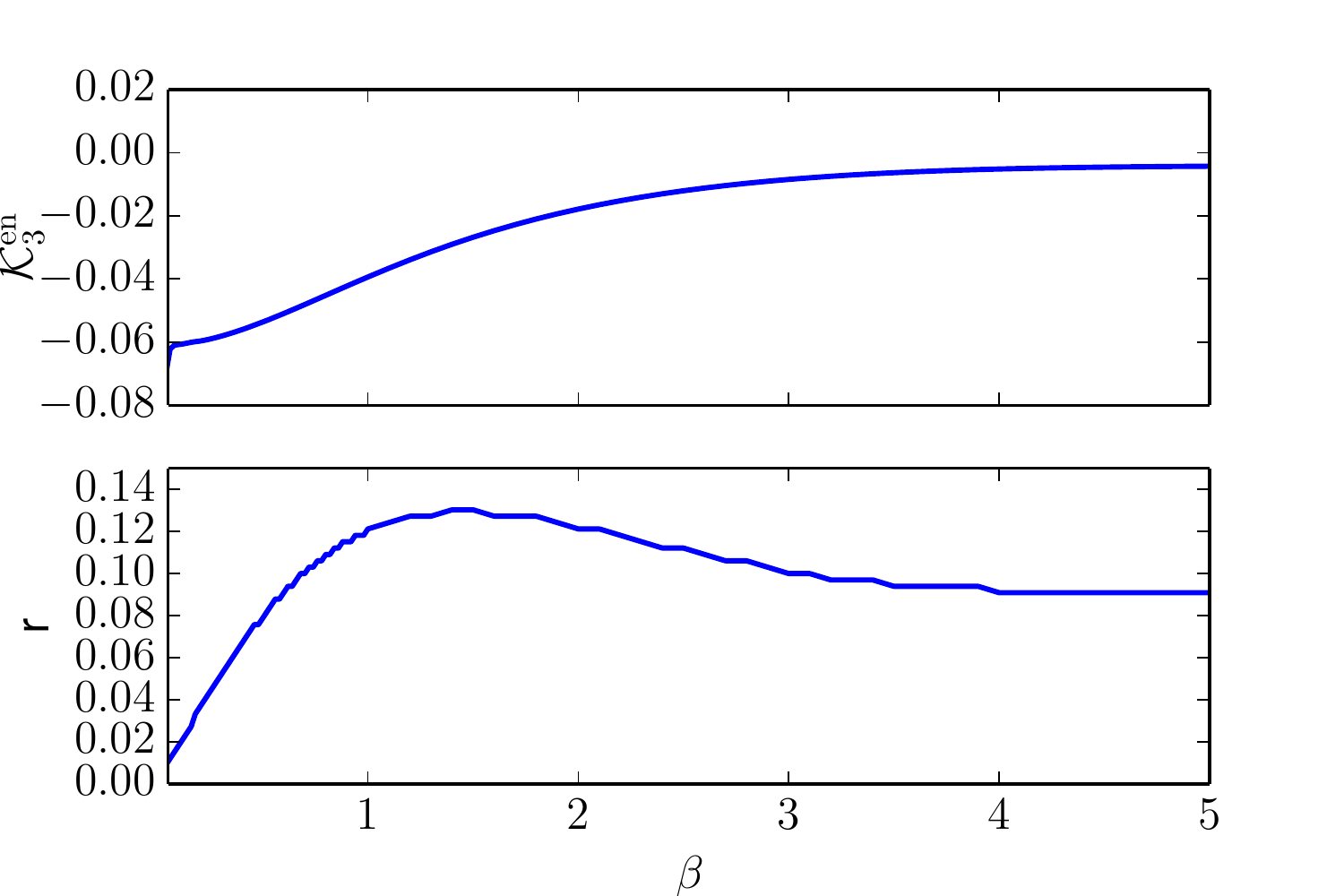}
    \caption{Maximal Leggett-Garg violation.
             Top: The smallest value of $\mathcal{K}^\text{en}_3 $
             as a function of $\beta$.
             Bottom: The corresponding values of $r$, where
             $\mathcal{K}^\text{en}_3 $ assumes the minimal value.
}
     \label{fig:LG_squeeze_beta}
\end{figure}

\section{Conclusion}
\label{sec:Conclusion}

We have shown that quantum work,
defined according to Eq.~\eqref{def:work} may show
statistical correlations that cannot be described by classical
macro-realism. 
This follows from the violation of Leggett-Garg
and entropic Leggett-Garg inequalities for
work measurements.
In a driven
two-level system as well as in a harmonic oscillator subject to squeezing,
these inequalities are violated for certain driving parameter
and initial temperatures. When both can be evaluated, the entropic and the normal Leggett-Garg 
inequalities do not necessarily identify the same correlations as non-classical, as their violation is only a sufficient but not a necessary 
criterion to abandon macro-realism. 
This points to the interest in developing 
tighter bounds to rule out the violation of macroscopic realism
over broader parameter ranges. 
\\
The study of temporal correlations and their implication for both foundational and practical questions resembles the situation in the 1950'es, 
where a multitude of optical phenomena could be described by stochastically fluctuating
classical fields, but where the Hanbury-Brown and Twiss measurements of (classical) intensity correlations spurred \cite{Glauber06} discussions about the 
general validity of classical modelling. This led to the insight that temporal fluctuations in intensity measurements can, indeed, exclude classical 
descriptions of the light field, and it stimulated the emergence of quantum optics as a research field. Non-classical properties of light are, e.g., witnessed by temporal 
noise correlations that violate Cauchy-Schwarz inequalities, anti-bunching, and higher-order interference effects, which have in several cases turned out to be useful properties, e.g., for precision sensing.
\\
In this spirit, our work is an  attempt to quantify temporal quantum correlations involved in thermodynamic processes and might be relevant 
for the evaluation and design of work extraction protocols \cite{Kammerlander16, Korzekwa16}
and (measurement based) quantum thermal machines \cite{Hayashi15}, where it was shown recently that the efficiency of cyclic processes 
may non-trivially involve correlations between subsequent cycles \cite{Watanabe17}.

\appendix
\allowdisplaybreaks

\section{Analytical formula for $G_{mn}(z)$}
\label{app:anal}

In this appendix, we show the explicit expression for the
matrix element $G_{mn}(z)$ derived in \cite{Satya85,Nieto97,Kim89}.
It yields
\begin{widetext}
  \begin{equation}
    G_{mn}(z)=\begin{cases}
                (-1)^{m/2} \frac{\sqrt{m! n!}}{\cosh r}
                 \sum_{i=1}^{N}
                     \frac{(-4)^i(\sinh z) ^{(m+n)/2 -2 i }  (2\cosh z )^{-(n+m)/2}}
                          {2 i! (1/2m - i)! (1/2n-i)!}
                   \qquad &\text{for} \ m,n \ \text{even} \\[2ex]
                    (-1)^{(m-1)/2} \frac{\sqrt{m! n!}}{\cosh r}
                 \sum_{i=1}^{N}
                     \frac{(-4)^i(\sinh z) ^{(m+n)/2 -2 i-1 }  (2\cosh z )^{-(n+m)/2-1}}
                          {(2 i+1)! (1/2(m-1) - i)! (1/2(n-1)-i)!}
                   \qquad &\text{for} \ m,n \ \text{odd}
       \\[2ex]
                     0  \qquad &\text{else}

              \end{cases}
\end{equation}

\end{widetext}

where the summation ends at $N= \min\{m/2,n/2\}$ for $n,m$ even
and $N= \min\{(m-1)/2,(n-1)/2\}$ for $n,m$ odd, respectively.

\

\begin{acknowledgments}
  The authors acknowledge financial support from the Villum foundation.
\end{acknowledgments}

\bibliography{literature}

\end{document}